\documentclass[twocolumn]{apex}
\usepackage{txfonts}

\usepackage{graphicx} \usepackage{bm} \usepackage{amsmath}
\usepackage{amssymb}

\title{Minimization of the Switching Time of a Synthetic Free Layer in Thermally Assisted Spin Torque Switching}

\author{
  Tomohiro Taniguchi and Hiroshi Imamura} 
\inst{
  Nanosystem Research Institute, AIST, 1-1-1 Umezono, Tsukuba 305-8568, Japan}

\abst{
  We theoretically studied the thermally assisted spin torque switching
  of a synthetic free layer and showed that the switching time is minimized if 
  the condition $H_{J}=|H_{\rm s}|/(2\alpha)$ is satisfied, 
  where $H_{J}$, $H_{\rm s}$, and $\alpha$ are 
  the coupling field of two ferromagnetic layers, 
  the amplitude of the spin torque, 
  and the Gilbert damping constant, respectively. 
  We also showed that 
  the coupling field of the synthetic free layer can be determined
  from the resonance frequencies of 
  the spin-torque diode effect. 
  }

\begin{document}
\maketitle


Spin random access memory (Spin RAM) using the tunneling
magnetoresistance (TMR) effect
\cite{yuasa04,parkin04} and spin torque switching
\cite{slonczewski89,berger96} is 
one of the important spin-electronics devices 
for future nanotechnology.  
For Spin RAM application, 
it is highly desired to realize the magnetic tunnel junction (MTJ) with 
high thermal stability $\Delta_{0}$, 
a low spin-torque switching current $I_{\rm c}$,
and a fast switching time.  
Recently, large thermal stabilities have been observed 
in anti-ferromagnetically \cite{hayakawa08} and
ferromagnetically \cite{yakata09} coupled synthetic free (SyF) layers
in MgO-based MTJs.  
In particular, the ferromagnetically coupled SyF layer is 
a remarkable structure 
because it shows thermal stability of more than 100 
with a low switching current \cite{yakata09}.


Since the coupling between the ferromagnetic layers in the SyF layer is 
indirect exchange coupling, 
we can systematically vary the sign and strength of the coupling field 
by changing the spacer thickness between 
the two ferromagnetic layers.  
As shown in ref. \cite{taniguchi11} , 
the thermal switching probability of the SyF layer is a 
double exponential function of the coupling field, 
and a tiny change in the coupling field can 
significantly increase or decrease the switching time. 
Therefore, it is of interest to physical science 
to study the dependence of the thermal switching time 
on the coupling field. 


In this paper, we theoretically studied the spin-current-induced
dynamics of magnetizations in an SyF layer of an MTJ.  
We found the optimum condition of the coupling field, 
which minimizes the thermally assisted spin torque switching time.
We showed that the coupling field of the
two ferromagnetic layers in the SyF layer can be determined
by using the spin torque diode effect.

\begin{figure}
\centerline{\includegraphics[width=0.8\columnwidth]{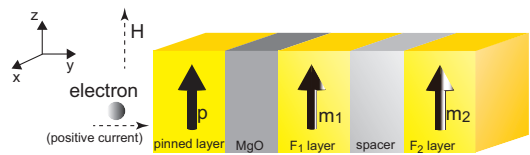}}
\caption{
         Schematic view of the SyF layer. 
         $\mathbf{m}_{k}$ and $\mathbf{p}$ are 
         the unit vectors pointing in the directions of 
         the magnetizations of the F${}_{k}$ and pinned layers, respectively. 
         The positive current is defined as 
         the electron flow from the pinned layer to the free layer. 
         $H$ represents the applied field. 
         \vspace{-3ex}}
\label{fig:fig1}
\end{figure}

Let us first briefly describe 
the thermal switching 
of the SyF layer 
in the weak coupling limit, 
$KV \gg JS$,
where $K$, $J$, $V$, and $S$ are 
the uniaxial anisotropy energy per unit volume, 
the coupling energy per unit area, 
and the volume and cross-sectional area 
of the single ferromagnetic layer, respectively. 
For simplicity, 
we assume that all the material parameters 
of the two ferromagnetic layers (F${}_{1}$ and F${}_{2}$) 
in the SyF layer are identical. 
A typical MTJ with an SyF layer is structured as 
a pinned layer/MgO barrier/ferromagnetic (F${}_{1}$) layer/nonmagnetic spacer/ferromagnetic (F${}_{2}$) layer
(see Fig. \ref{fig:fig1}), 
where the F${}_{1}$ and F${}_{2}$ layers are 
ferromagnetically coupled due to the interlayer exchange coupling \cite{yakata09}. 
The F${}_{1}$ and F${}_{2}$ layers have 
uniaxial anisotropy along the $z$ axis 
and two energy minima at $\mathbf{m}_{k}=\pm \mathbf{e}_{z}$, 
where $\mathbf{m}_{k}$ is the unit vector 
pointing in the direction of the magnetization of the F${}_{k}$ layer. 
The spin current injected from the pinned layer to the F${}_{1}$ layer 
exerts spin torque on the magnetization of the F${}_{1}$ layer \cite{comment1}. 
Then, the magnetization of the F${}_{1}$ layer 
switches its direction due to the spin torque, 
after which 
the magnetization of the F${}_{2}$ layer switches its direction 
due to coupling. 
By increasing the coupling field, 
the potential height of the F${}_{1}$ (F${}_{2}$) layer 
for the switching becomes high (low), 
which makes the switching time 
of the F${}_{1}$ (F${}_{2}$) layer long (short). 
Then, a minimum of the total switching time 
appears at a certain coupling field, 
as we shall show below. 


The switching probability from the parallel (P) to antiparallel (AP) alignment 
of the pinned and free layer magnetizations is given by \cite{taniguchi11} 
\begin{equation}
  P
  =
  1
  -
  (\nu_{\rm F_{1}}{\rm e}^{-\nu_{\rm F_{2}}t}-\nu_{\rm F_{2}}{\rm e}^{-\nu_{\rm F_{1}}t})/(\nu_{\rm F_{1}}-\nu_{\rm F_{2}}),
  \label{eq:switching_rate}
\end{equation}
where $\nu_{{\rm F}_{k}} \!=\! f_{{\rm F}_{k}}\exp(-\Delta_{{\rm F}_{k}})$ is 
the switching rate of the F${}_{k}$ layer. 
The attempt frequency is given by 
$f_{{\rm F}_{k}} \!=\! f_{0}\delta_{k}$,
where 
$f_{0}
  \!=\! 
  [\alpha\gamma H_{\rm an}/(1 \!+\! \alpha^{2})]
  \sqrt{\Delta_{0}/\pi}$, 
$\delta_{1}
  \!=\! 
  [1 \!-\! (H \!+\! H_{J} \!+\! H_{\rm s}/\alpha)^{2}/H_{\rm an}^{2}]
  [1 \!+\! (H \!+\! H_{J} \!+\! H_{\rm s}/\alpha)/H_{\rm an}]$,
and 
$\delta_{2}
  \!=\! 
  [1 \!-\! (H \!-\! H_{J})^{2}/H_{\rm an}^{2}]
  [1 \!+\! (H \!-\! H_{J})/H_{\rm an}]$. 
$\alpha$, $\gamma$, $H$, $H_{\rm an} \!=\! 2K/M$, $H_{J} \!=\! J/(Md)$, and $\Delta_{0} \!=\! KV/(k_{\rm B}T)$ are 
the Gilbert damping constant, 
gyromagnetic ratio, 
applied field, 
uniaxial anisotropy field, 
coupling field, 
and thermal stability, respectively, 
and $d$ is the ferromagnetic layer thickness. 
$\Delta_{{\rm F}_{k}}$ is given by \cite{taniguchi11,comment2}
\begin{equation}
  \Delta_{\rm F_{1}}
  =
  \Delta_{0}
  \left[
    1
    +
    (H+H_{J}+H_{\rm s}/\alpha)/H_{\rm an}
  \right]^{2},
  \label{eq:Delta_F1}
\end{equation}
\begin{equation}
  \Delta_{\rm F_{2}}
  =
  \Delta_{0}
  \left[
    1
    +
    (H-H_{J})/H_{\rm an}
  \right]^{2}.
  \label{eq:Delta_F2}
\end{equation}
$\Delta_{\rm F_{1}}$ is the potential height of the F${}_{1}$ layer
before the F${}_{2}$ layer switches its magnetization 
while $\Delta_{\rm F_{2}}$ is the potential height of the F${}_{2}$ layer 
after the F${}_{1}$ layer switches its magnetization. 
$H_{\rm s}=\hbar \eta I/(2eMSd)$ is
the amplitude of the spin torque 
in the unit of the magnetic field, 
where $\eta$ is the spin polarization of the current $I$. 
The positive current corresponds to 
the electron flow from the pinned to the F${}_{1}$ layer; 
i.e., the negative current $I$ ($H_{\rm s}<0$) 
induces the switching of the F${}_{1}$ layer. 
The field strengths should satisfy 
$|H \!+\! H_{J} \!+\! H_{\rm s}/\alpha|/H_{\rm an} \!<\! 1$ 
and $|H \!-\! H_{J}|/H_{\rm an} \!<\! 1$ 
because eq. (\ref{eq:switching_rate}) is valid 
in the thermal switching region. 
In particular,  $|H \!+\! H_{J} \!+\! H_{\rm s}/\alpha|/H_{\rm an} \!<\! 1$ means that 
$|I|<|I_{\rm c}|$. 
The effect of the field like torque is neglected in Eq. (\ref{eq:Delta_F1}) 
because its magnitude, $\beta H_{\rm s}$ where the beta term satisfies $\beta < 1$, 
is less than 1 Oe in the thermal switching region 
and thus, negligible. 

\begin{figure}
\centerline{\includegraphics[width=0.7\columnwidth]{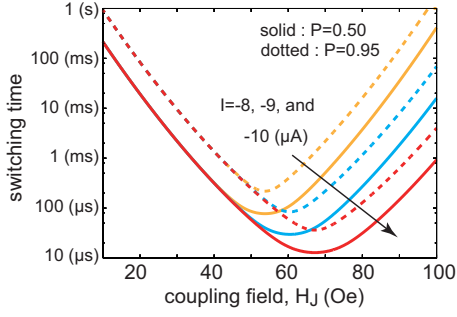}}
\caption{
         Dependences of the switching time at 
         $P=0.50$ (solid lines) and $P=0.95$ (dotted lines) 
         on the coupling field $H_{J}$
         with currents 
         $I=-8$ (yellow), $-9$ (blue), and $-10$ (red) $\mu$A. 
         \vspace{-3ex}}
\label{fig:fig2}
\end{figure}

Figure \ref{fig:fig2} shows
the dependences of the switching times 
at $P=0.50$ and $P=0.95$ 
on the coupling field 
with the currents 
(a) $-8$, (b) $-9$, and 
(c) $-10$ $\mu$A. 
The values of the parameters are taken to be 
$\alpha \!=\! 0.007$, 
$\gamma \!=\! 17.32$ MHz/Oe, 
$H_{\rm an} \!=\! 200$ Oe, 
$M \!=\! 995$ emu/c.c., 
$S \!=\! \pi \!\times\! 80 \!\times\! 35$ nm${}^{2}$, 
$d \!=\! 2$ nm, 
and $T \!=\! 300$ K \cite{yakata09}.
The values of $H$ and $\eta$ are taken to be 
$-65$ Oe and $0.5$, respectively. 
The value of $H$ is chosen so as to make 
the potential heights for the switching low as much as possible 
($|H \!+\! H_{J} \!+\! H_{\rm s}/\alpha|/H_{\rm an} \!\lesssim\! 1$ 
and $|H \!-\! H_{J}|/H_{\rm an} \!\lesssim\! 1$). 
As shown in Fig. \ref{fig:fig2}, 
the switching time is minimized
at a certain coupling field. 
We call this $H_{J}$ as the optimum coupling field 
for the fast thermally assisted spin torque switching. 


Let us estimate the optimum coupling field. 
For a small $H_{J}$, 
the switching time of the F${}_{2}$ layer is the main determinant of
the total switching time; 
thus, eq. (\ref{eq:switching_rate}) can be approximated as 
$P \!\simeq\! 1 \!-\! {\rm e}^{-\nu_{\rm F_{2}}t}$. 
By increasing $H_{J}$, 
$\nu_{\rm F_{2}}$ increases 
and the switching time ($\!\sim\! 1/\nu_{\rm F_{2}}$) decreases. 
Fast switching is achieved 
for $\nu_{{\rm F}_{2}} \!\sim\! \nu_{{\rm F}_{1}}$ 
in this region. 
On the other hand, 
for a large $H_{J}$, 
the switching time of the F${}_{1}$ layer dominates, 
and eq. (\ref{eq:switching_rate}) is approximated as 
$P \!\simeq\! 1 \!-\! {\rm e}^{-\nu_{\rm F_{1}}t}$. 
The switching time ($\!\sim\! 1/\nu_{\rm F_{1}}$) decreases 
with decreasing $H_{J}$. 
Fast switching in this region is also achieved 
for $\nu_{{\rm F}_{1}} \!\sim\! \nu_{{\rm F}_{2}}$.
The switching rate $\nu_{{\rm F}_{k}}$ is mainly determined by 
$\Delta_{{\rm F}_{k}}$. 
By putting $\Delta_{\rm F_{1}} \!=\! \Delta_{\rm F_{2}}$, 
the optimum coupling field is obtained as
\begin{equation}
  H_{J}
  =
  |H_{\rm s}|/(2\alpha).
  \label{eq:optimized_coupling}
\end{equation}
This is the main result of this paper. 
The values obtained with eq. (\ref{eq:optimized_coupling}) 
for $I \!=\! -8, -9$ and $-10$ $\mu$A 
are 53.7, 60.5, and 67.2 Oe, respectively, 
which show good agreement
with Fig. \ref{fig:fig2}. 


The condition $\nu_{\rm F_{1}} \!\simeq\! \nu_{\rm F_{2}}$ means that 
the most efficient switching can be realized 
when two switching processes of the F${}_{1}$ and F${}_{2}$ layers occur with the same rate. 
$\nu_{\rm F_{1}}>\nu_{\rm F_{2}}$ means that 
the magnetization of the F${}_{1}$ layer can easily switch 
due to a large spin torque. 
However, the system should stay in this state for a long time 
because of a small switching rate of the F${}_{2}$ layer. 
On the other hand, 
when $\nu_{\rm F_{1}}<\nu_{\rm F_{2}}$, 
it takes a long time to switch the magnetization of the F${}_{1}$ layer. 
Thus, when $\nu_{\rm F_{1}}$ and $\nu_{\rm F_{2}}$ are different, 
the system stays in an unswitched state of the F${}_{1}$ or F${}_{2}$ layer 
for a long time, 
and the total switching time becomes long. 
For thermally assisted field switching, 
we cannot find the optimum condition of the switching time 
because the switching probabilities of 
the F${}_{1}$ and F${}_{2}$ layers are the same.
Factor 2 in eq. (\ref{eq:optimized_coupling}) arises from
the fact that 
$H_{J}$ affects the switchings of 
both the F${}_{1}$ and F${}_{2}$ layers, 
while $H_{\rm s}$ assists that of only the F${}_{1}$ layer. 
When $H_{J}\ll |H_{\rm s}|/(2\alpha)$, 
the total switching time 
is independent of the current strength, 
because the total switching time in this region 
is mainly determined by 
the switching time of the F${}_{2}$ layer, 
which is independent of the current. 
In the strong coupling limit, 
$KV \ll JS$, 
two magnetizations switch simultaneously \cite{taniguchi11}, 
and the switching time is independent of the coupling field. 


For the AP-to-P switching, 
the factors $\delta_{k}$ and $\Delta_{{\rm F}_{k}}$ are given by 
$\delta_{1}
  \!=\! 
  [1 \!-\! (H \!-\! H_{J} \!+\! H_{\rm s}/\alpha)^{2}/H_{\rm an}^{2}]
  [1 \!-\! (H \!-\! H_{J} \!+\! H_{\rm s}/\alpha)/H_{\rm an}]$,
$\delta_{2}
  \!=\! 
  [1 \!-\! (H \!+\! H_{J})^{2}/H_{\rm an}^{2}]
  [1 \!-\! (H \!+\! H_{J})/H_{\rm an}]$,
$\Delta_{{\rm F}_{1}}
  \!=\!
  \Delta_{0}
  \left[1 \!-\! (H \!-\! H_{J} \!+\! H_{\rm s}/\alpha)/H_{\rm an}\right]^{2}$, 
and 
$\Delta_{\rm F_{2}}
  \!=\! 
  \Delta_{0}
  \left[1 \!-\! (H \!+\! H_{J})/H_{\rm an}\right]^{2}$. 
In this case, 
a positive current ($H_{\rm s} \!>\! 0$) induces the switching. 
By setting $\Delta_{\rm F_{1}} \!=\! \Delta_{\rm F_{2}}$, 
the optimum coupling field is obtained as
$H_{J} \!=\! H_{\rm s}/(2\alpha)$. 
Thus, for both P-to-AP and AP-to-P switchings, 
the optimum coupling field is expressed as 
$H_{J} \!=\! |H_{\rm s}|/(2\alpha)$. 


In the case of the anti-ferromagnetically coupled SyF layer, 
$H \!+\! H_{J}$ and $H \!-\! H_{J}$ in 
eqs. (\ref{eq:Delta_F1}) and (\ref{eq:Delta_F2}) 
should be replaced by 
$H \!+\! |H_{J}|$ and 
$\!-\! H \!-\! |H_{J}|$, respectively, 
where the sign of the coupling field is negative ($H_{J}<0$). 
The optimum condition is given by 
$|H_{J}| \!=\! -H \!+\! |H_{\rm s}|/(2\alpha)$, 
where the negative current is assumed to 
enhance the switching of the F${}_{1}$ layer. 
For a sufficiently large positive field 
$H \!>\! |H_{\rm s}|/(2\alpha)$, 
this condition 
cannot be satisfied 
because $\nu_{\rm F_{1}}$ is always smaller than $\nu_{\rm F_{2}}$. 


One might notice that 
the condition 
$\Delta_{\rm F_{1}} \!=\! \Delta_{\rm F_{2}}$ for the ferromagnetically coupled SyF layer 
has another solution 
$|H_{\rm s}|/(2\alpha) \!=\! H \!+\! H_{\rm an}$, 
which is independent of the coupling field. 
We exclude this solution 
because such $H$ and $H_{\rm s}$ cannot satisfy 
the conditions for the thermal switching regions 
$|H \!+\! H_{J} \!+\! H_{\rm s}/\alpha| \!<\! H_{\rm an}$ and 
$|H \!-\! H_{J}| \!<\! H_{\rm an}$ simultaneously. 
Similarly, 
for the anti-ferromagnetically coupled SyF layer, 
we exclude the solution 
$|H_{\rm s}|/(2\alpha) \!=\! H_{\rm an}$
obtained from $\Delta_{\rm F_{1}} \!=\! \Delta_{\rm F_{2}}$.


The natural question from the above discussion is 
how large the coupling field is.  
The coupling field of a large plane film can be determined
from two ferromagnetic resonance (FMR) frequencies
\cite{zhang94,lindner05} 
corresponding to the acoustic and optical modes, 
which depend on $H_{J}$.  
The antiferromagnetic coupling field can also be
determined by the magnetization curve \cite{hayakawa08}, 
in which finite magnetization appears when the applied field exceeds the
saturation field $H_{\rm s} \!=\! -2H_{J}$.  
These methods are, however, not
applicable to nanostructured ferromagnets such as the Spin RAM cells
because 
the signal intensity is proportional to 
the volume of the ferromagnet, 
and thus,
the intensity from the Spin RAM cell is negligibly small.  
It is desirable to measure 
the coupling field of each cell 
because $H_{J}$ strongly depends on the surface state 
and may differ significantly 
among the cells obtained from a single film plane.


Here, we propose that 
the coupling field 
can be determined 
by using the spin torque diode effect \cite{tulapurkar05,kubota08,sankey08}
of the SyF layer. 
This method is applicable to a nanostructured ferromagnet, 
although the basic idea is similar to that of FMR measurement. 


The spin torque diode effect is measured 
by applying an alternating current $I_{\rm a.c.}\cos(2\pi ft)$ to an MTJ, 
which induces oscillating spin torque 
on the magnetization of the F${}_{1}$ layer. 
The free layer magnetizations 
oscillate due to the oscillating spin torque and the coupling, 
which lead to the oscillation of 
the TMR 
$R_{\rm TMR} 
  \!=\! 
  R_{\rm P}
  \!+\! 
  (1 \!-\! \mathbf{p}\cdot\mathbf{m}_{1})\Delta R/2$ 
and the d.c. voltage $V_{\rm d.c.}$. 
Here, $\Delta R \!=\! R_{\rm AP} \!-\! R_{\rm P}$, and 
$R_{\rm P}$ and $R_{\rm AP}$ correspond to 
the resistances at the parallel and antiparallel alignments 
of $\mathbf{p}$ and $\mathbf{m}_{1}$, respectively. 
$\mathbf{p}$ is the unit vector pointing in 
the direction of the pinned layer magnetization. 
$V_{\rm d.c.}$ is given by 
\begin{equation}
  V_{\rm d.c.}
  =
  \frac{1}{T}
  \int_{0}^{T} {\rm d}t 
  I_{\rm a.c.}
  \cos(2\pi ft)
  \frac{-\Delta R}{2}
  \mathbf{p}
  \cdot
  \mathbf{m}_{1},
  \label{eq:voltage}
\end{equation}
where $T \!=\! 1/f$. 
The SyF layer shows large peaks of d.c. voltage 
at the FMR frequencies of 
the acoustic $f_{\rm acoustic}$ and optical $f_{\rm optical}$ modes. 
The coupling field can be determined
from these frequencies. 


\begin{figure}
\centerline{
\includegraphics[width=0.6\columnwidth]{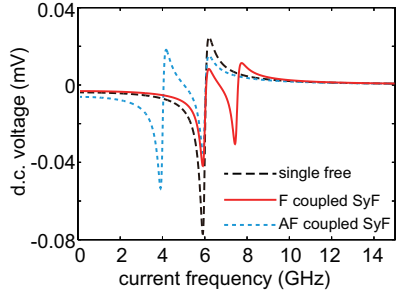}}
\caption{Dependences of the spin torque diode voltage 
         of the single free layer (solid), 
         the ferromagnetically (F) coupled SyF layer (dotted), 
         and the anti-ferromagnetically (AF) coupled SyF layer (dashed)
         on the applied current frequency. 
         \vspace{-3ex}}
\label{fig:fig3}
\end{figure}


The resonance frequency of 
the ferromagnetically coupled system 
is obtained as follows. 
The free energy of the SyF layer 
is given by 
\begin{equation}
\begin{split}
  \frac{F}{MV}
  =&
  -\mathbf{H}
  \cdot
  \left(
    \mathbf{m}_{1}
    +
    \mathbf{m}_{2}
  \right)
  -
  \frac{H_{\rm an}}{2}
  \left[
    (\mathbf{m}_{1}\cdot\mathbf{e}_{z})^{2}
    +
    (\mathbf{m}_{2}\cdot\mathbf{e}_{z})^{2}
  \right]
\\
  &+
  2\pi M
  \left[
    (\mathbf{m}_{1}\cdot\mathbf{e}_{y})^{2}
    +
    (\mathbf{m}_{2}\cdot\mathbf{e}_{y})^{2}
  \right]
  -
  H_{J}
  \mathbf{m}_{1}
  \cdot
  \mathbf{m}_{2},
\end{split}
\end{equation}
where the first, second, third, and fourth terms are
the Zeeman energy, 
uniaxial anisotropy energy, 
demagnetization field energy, 
and coupling energy, respectively. 
The $y$ and $z$ axes are 
normal to the plane 
and parallel to the easy axis, respectively. 
The applied field, 
$\mathbf{H}
  \!=\! 
  H(\sin\theta_{H}\mathbf{e}_{x} \!+\! \cos\theta_{H}\mathbf{e}_{z})$, 
lies in the $xz$ plane with angle $\theta_{H}$ from the $z$ axis. 
The equilibrium point 
is located at 
$\mathbf{m}_{1} 
  \!=\! 
  \mathbf{m}_{2}
  \!=\!
  \mathbf{m}^{(0)} 
  \!=\! 
  (\sin\theta_{0},0,\cos\theta_{0})$, 
where $\theta_{0}$ satisfies 
$H \sin(\theta_{0} \!-\! \theta_{H}) 
  \!+\! 
  H_{\rm an}\sin\theta_{0}\cos\theta_{0}
  \!=\! 
  0$.
We employ a new $XYZ$ coordinate 
in which the $Y$ and $Z$ axes are parallel to 
the $y$ axis and $\mathbf{m}^{(0)}$, respectively,
and denote a small component of the magnetization 
around $\mathbf{m}^{(0)}$ as 
$\delta \mathbf{m}_{k} \!=\! (m_{kX},m_{kY},0)$. 
The magnetization dynamics is desribed by using 
the Landau-Lifshitz-Gilbert (LLG) equation 
${\rm d}\mathbf{m}_{k}/{\rm d}t \!=\! 
  -\gamma \mathbf{m}_{k} \!\times\! \mathbf{H}_{k}
  \!+\! 
  \alpha \mathbf{m}_{k} \!\times\! ({\rm d}\mathbf{m}_{k}/{\rm d}t)$, 
where $\mathbf{H}_{k} \!=\! -(MV)^{-1}\partial F/\partial\mathbf{m}_{k}$ 
is the field acting on $\mathbf{m}_{k}$. 
By assuming the oscillating solution ($\propto {\rm e}^{2\pi{\rm i}\tilde{f} t}$)
of $m_{kX}$ and $m_{kY}$, 
keeping the first-order terms of $m_{kX}$ and $m_{kY}$, 
and neglecting the damping term, 
the LLG equations can be 
linearized as $\mathsf{M}(m_{1X},m_{1Y},m_{2X},m_{2Y})^{\rm t} \!=\! 0$. 
The nonzero components of the coefficient matrix are 
$\mathsf{M}_{11} \!=\! \mathsf{M}_{22} \!=\! \mathsf{M}_{33} \!=\! \mathsf{M}_{44} \!=\! 2\pi{\rm i}\tilde{f}/\gamma$, 
$\mathsf{M}_{12} 
  \!=\! 
  \mathsf{M}_{34} 
  \!=\! 
  [H \cos(\theta_{H} \!-\! \theta_{0}) \!+\! H_{\rm an}\cos^{2}\theta_{0} \!+\! H_{J} \!+\! 4\pi M]$, 
$\mathsf{M}_{21}
  \!=\! 
  \mathsf{M}_{43}
  \!=\!
  -[H \cos(\theta_{H} \!-\! \theta_{0}) \!+\! H_{\rm an}\cos 2 \theta_{0} \!+\! H_{J}]$, 
and 
$\mathsf{M}_{14}
  \!=\!
  -\mathsf{M}_{23}
  \!=\!
  \mathsf{M}_{32}
  \!=\!
  -\mathsf{M}_{41}
  \!=\! 
  -H_{J}$. 
The FMR resonance frequencies are obtained
under the condition ${\rm det}[\mathsf{M}] \!=\! 0$, 
and are given by 
$f_{\rm acoustic} \!=\! \gamma\sqrt{h_{1}h_{2}}/(2\pi)$ 
and 
$f_{\rm optical} \!=\! \gamma\sqrt{(h_{1} \!+\! 2H_{J})(h_{2} \!+\! 2H_{J})}/(2\pi)$, 
where 
$h_{1} 
  \!=\!
  H \cos(\theta_{H} \!-\! \theta_{0})
  \!+\!
  H_{\rm an} \cos 2 \theta_{0}$ 
and 
$h_{2}
  \!=\!
  H \cos(\theta_{H}-\theta_{0})
  \!+\!
  H_{\rm an} \cos^{2}\theta_{0}
  \!+\!
  4\pi M$. 
$H_{J}$ can be determined
from these frequencies.
For the anti-ferromagnetically coupled system,  
$\mathbf{m}_{1} \neq \mathbf{m}_{2}$ in equilibrium in general, 
and the resonance frequencies are obtained  
by solving the $4 \!\times\! 4$ matrix equation.


Figure \ref{fig:fig3} shows the dependences 
of the d.c. voltage $V_{\rm d.c.}$ 
of the single free layer (solid) 
and the ferromagnetically (dotted) and anti-ferromagnetically
(dashed) coupled SyF layers 
on the applied current frequency 
calculated by solving the LLG equations
of the F${}_{1}$, F${}_{2}$, and pinned layers. 
The spin torque term, 
$\gamma 
  H_{\rm s} 
  \mathbf{m}_{1} 
  \!\times\! 
  (\mathbf{p} \!\times\! \mathbf{m}_{1}) 
  \!+\! 
  \gamma 
  \beta 
  H_{\rm s} 
  \mathbf{p} 
  \!\times\! 
  \mathbf{m}_{1}$, 
is added to the LLG equation of the F${}_{1}$ layer. 
Here the field like torque is taken into account 
because it affects the shape of $V_{\rm d.c.}$ significantly \cite{tulapurkar05}. 
The magnetic field acting on $\mathbf{p}$ is given by 
$\mathbf{H}_{\rm pin} 
  \!=\!
  \mathbf{H}
  \!-\! 
  4\pi Mp_{y}\mathbf{e}_{y}
  \!+\!
  (H_{\rm an}p_{z} \!+\! H_{\rm p}) \mathbf{e}_{z}$, 
where $H_{\rm p}$ is the pinning field 
due to the bottom anti-ferromagnetic layer \cite{yakata09}. 
In Fig. \ref{fig:fig3}, 
$I_{\rm a.c.}=0.1$ mA, 
$\Delta R=400$ $\Omega$, 
$H \!=\! 200$ Oe, 
$|H_{J}| \!=\! 100$ Oe, 
$H_{\rm p} \!=\! 2$ kOe, 
$\theta_{H} \!=\! 30^{\circ}$ 
and $\beta \!=\! 0.3$ \cite{tulapurkar05}. 
The $f_{\rm acoustic}$ and $f_{\rm optical}$ 
of the ferromagnetically coupled SyF layer 
are estimated to be 5.98 and 7.50 GHz, respectively, 
which show good agreement 
with the peak points in Fig. \ref{fig:fig3}. 
These results indicate that 
the spin torque diode effect is useful 
in determining the coupling field. 


In summary, 
we theoretically studied the dependence of 
the thermally assisted spin torque switching time 
of a SyF layer on the coupling field. 
We found that the switching time is minimized if the 
condition of $H_{J}=|H_{\rm s}|/(2\alpha)$ is satisfied. 
We showed that the coupling field can be determined
from the resonance frequency of the spin torque diode effect. 


The authors would like to acknowledge 
H. Kubota, 
T. Saruya, 
D. Bang, 
T. Yorozu, 
H. Maehara, 
and S. Yuasa
of AIST 
for their support 
and the discussions they had with us.


\begin{thebibliography}{20}
\bibitem{comment2}
In ref. \cite{taniguchi11} , 
$\Delta_{\rm F_{1}}$ is expressed as 
$\Delta_{0}[1+(H_{\rm appl}+H_{J})/H_{\rm an}]^{2}(1-I/I_{\rm c})^{2}$, 
which is equivalent to eq. (2).
\end{thebibliography}
\end{document}